\documentstyle[prd,aps]{revtex}

\newcommand{\bea}{\begin{eqnarray}}
\newcommand{\eea}{\end{eqnarray}}

\begin{document}

%%%%%%%%%%%%%%%%%%%%%%%%%%%%%%%%%%%%%%%%%%%%%%%%%%%%%%%%%%%%%%%
\draft
%  For 2 column format.
\twocolumn[\hsize\textwidth\columnwidth\hsize\csname
@twocolumnfalse\endcsname

%%%%%%%%%%%%%%%%%%%%%%%%%%%%%%%%%%%%%%%%%%%%%%%%%%%%%%%%%%%%%%%
\title{Density Spectrums from Kinetic Inflations}
\author{Jai-chan Hwang}
\address{Department of Astronomy and Atmospheric Sciences,
         Kyungpook National University, Taegu, Korea}
\author{Hyerim Noh}
\address{Korea Astronomy Observatory,
         San 36-1, Whaam-dong, Yusung-gu, Daejon, Korea}
\date{\today}
\maketitle

%%%%%%%%%%%%%%%%%%%%%%%%%%%%%%%%%%%%%%%%%%%%%%%%%%%%%%%%%%%%%%%
\begin{abstract}

The pole-like accelerated expansion stages purely driven by the coupling 
between the gravity and the dilaton field without referring to the 
potential term can be realized in a class of generalized gravity theories.
We consider three such scenarios based on the scalar-tensor gravity, 
the induced gravity and the string theory.
Quantum fluctuations during the expansion stages (including more general
situations) can be derived in exact analytic forms.
Assuming that the pole-like acceleration stage provides a viable inflation 
scenario in the early universe we derive the generated classical density 
spectrums.
The generated classical density field shows a generic tilted spectrum with 
$n \simeq 4$ which differs from the observed spectrum supporting $n \simeq 1$.

\end{abstract}

\noindent

%%%%%%%%%%%%%%%%%%%%%%%%%%%%%%%%%%%%%%%%%%%%%%%%%%%%%%%%%%%%%%%
%  For 2 column format.
\vskip2pc]
%%%%%%%%%%%%%%%%%%%%%%%%%%%%%%%%%%%%%%%%%%%%%%%%%%%%%%%%%%%%%%%

%%%%%%%%%%%%%%%%%%%%%%%%%%%%%%%%%%%%%%%%%%%%%%%%%%%%%%%%%%%%%%%
\section{Introduction}
                                \label{sec:Introduction}

A pole-like accelerated expansion stage can be realized by 
a nontrivial coupling between the dilaton field and the gravity 
without referring to the potential term.
These types of accelerated expansion can be considered as the potential 
candidates for inflation models \cite{Inflation}.
One scenario based on the induced gravity is proposed by the authors 
of \cite{Pollock}.
Recently, the string theory has motivated an action which is 
also known to allow a similar acceleration stage \cite{pre-big-bang}.
The similar inflation scenario in the context of the scalar-tensor gravity 
is proposed in \cite{Levin} and is termed as the kinetic inflation.
We adopt the name ``kinetic inflation'' for the pole-like 
acceleration stages realized in various generalized gravity without 
the potential term.
We know that all these theories are related with each other by
simple field redefinitions or conformal transformations.
Although the acceleration stages can be realized, the specific realization of 
inflation model which resolves the conventional cosmological flatness 
and horizon problems with a successful graceful exit is yet to be made,
and is currently under active investigation \cite{pre-big-bang,Levin}.

During the expansion stage the quantum fluctuations in the dilaton field
and the metric may naturally arise from the vacuum expectation values
of the fluctuating field and the metric.
In this paper we will present the quantum fluctuations generated during
the expansion stages in analytic forms for the scalar type perturbations.
We will show that the generated power spectrums have the same spectral index 
independently of the specifics of considered gravity or the expansion stage.

Assuming that the acceleration stage provides a viable inflation stage
in the early universe we can derive the generated spectrums of the 
large scale density field in the second horizon crossing epoch in 
the conventional era of cosmological evolution.
In this part we will {\it consider} a scenario where a generalized gravity
theory has the roles of the gravity theory in the early universe and 
Einstein's theory takes over the role of gravity at some point.
That is, the kinetic inflation is {\it accepted} as the acceleration stage
while the observationally relevant scales leave the horizon scales.
In such a case, the classical fluctuations in the large scale arise 
from quantum fluctuations of the metric and the scalar field during 
the kinetic polar inflation stage.
Assuming that the fluctuations become superhorizon scale and are classicalized,
it is known that the fluctuations freeze independently of the changes in the 
background equation of state and also changes in the underlying gravity sector.
Such a freezing is conveniently characterized by a (temporally) conserved 
quantity which is usually represented by a curvature fluctuation in a certain 
gauge choice (spatial characters of the fluctuations are always conserved
during the linear evolution stage independently of the horizon scale;
in this sense, no structure formation [like self-organization]
arises in the linear theory). 
The observationally relevant spectrums we will derive in the following 
will be valid as long as the transition of the gravity sector occurs while the
observationally relevant scales stay in the superhorizon scale.

The generalized gravity theories to be considered in the following are 
simple subsets of the generalized gravity theories generally studied in 
\cite{GGT-H,GGT-QFT}.
Thus, we will present the structure generation and evolution processes
in the generalized gravity theories by reducing the general results in 
\cite{GGT-H,GGT-HN,GGT-QFT}.
Our study presented below will be based on the original frame of
the generalized gravity theory without referring to the 
conformal transformation.
Parallel analyses in the pre-big bang scenario based on the low energy
effective action of the string theory are presented in \cite{GGT-string}.

In Sec. \ref{sec:Formulation} we summarize the general and unified formulation
for handling the quantum generation and classical evolution processes
of the scalar type perturbations in the generalized gravity theories.
In Sec. \ref{sec:KI} applications are made to the expansion stages 
driven by the nontrivial coupling between the gravity and the dilaton field
without potential terms.
Cases include the scalar-tensor theory, the induced gravity, the string 
theory, and Einstein gravity with a minimally coupled scalar field.
We present the power spectrums of the quantum fluctuations in analytic forms.
Useful equations are deposited in the Appendix for a convenient reference.
We show that the spectrums are valid for various types of 
expansion stages allowed by the potential-less assumption.
We discuss the relations among results in different gravity theories.
In Sec. \ref{sec:MSF}, for comparison, we briefly summarize the results 
for the ordinary inflation based on the field potential in a minimally 
coupled scalar field. 
In Sec. \ref{sec:Spectrums} we derive the generated density spectrums in the
second horizon crossing epoch.
We assumed that the underlying gravity has swiched from the
generalized gravity to the Einstein one while the relevant perturbation
scales were in the superhorizon size.
Sec. \ref{sec:Discussion} is a brief discussion.
We set $c \equiv 1 \equiv 8 \pi G$.

%%%%%%%%%%%%%%%%%%%%%%%%%%%%%%%%%%%%%%%%%%%%%%%%%%%%%%%%%%%%%%
\section{General Formulation}
                                        \label{sec:Formulation}

In \cite{GGT-H,GGT-QFT,GGT-HN,GGT-CT} we have considered gravity theories
represented by the following action 
\bea
   S = \int d^4 x \sqrt{-g} \left[ {1 \over 2} f (\phi, R)
       - {1\over 2} \omega (\phi) \phi^{;a} \phi_{,a} - V(\phi) \right].
   \label{GGT-action}
\eea
The gravitational field equation and the equation of motion are:
\bea
   & & R_{ab} = {1\over F} \Bigg[
       \omega \phi_{,a} \phi_{,b}
       + g_{ab} \left( V + {RF - f \over 2} \right) 
   \nonumber \\
   & & \qquad \qquad
       + F_{,a;b} + {1 \over 2} g_{ab} \Box F \Bigg],
   \nonumber \\
   & & \Box \phi + {1 \over 2 \omega} \left(
       \omega_{,\phi} \phi^{;a} \phi_{,a}
       + f_{,\phi} - 2 V_{,\phi} \right) = 0,
   \label{GGT-GFE}
\eea
where $F \equiv \partial f/ \partial R$.

We consider a homogeneous, isotropic and flat cosmological model
with the general scalar type perturbations
\bea
   d s^2 
   &=& - \left( 1 + 2 \alpha \right) d t^2 - a \chi_{,\alpha} d t d x^\alpha
   \nonumber \\
   & & + \;
       a^2 \delta_{\alpha\beta} \left( 1 + 2 \varphi \right)
       d x^\alpha d x^\beta,
   \label{metric-general}
\eea
where $\alpha ({\bf x}, t)$, $\chi ({\bf x}, t)$ and $\varphi ({\bf x}, t)$
are the scalar type metric perturbations.
Without losing the generality, we have taken a spatial gauge choice which 
(in combination with the temporal gauge condition left for our freedom
to choose) will fix the spatial gauge mode completely.
We consider perturbations in the scalar (or dilaton) field as
\bea
   \phi ({\bf x}, t) = \bar \phi (t) + \delta \phi ({\bf x}, t),
\eea
where a background quantity is indicated by an overbar
which will be neglected unless necessary.
Equations for the background are:
\bea
   & & H^2 = {1 \over 3F} \left( {\omega \over 2} \dot \phi^2
       + V + {RF - f \over 2} - 3 H \dot F \right),
   \nonumber \\
   & & \ddot \phi + 3 H \dot \phi + {1\over 2 \omega}
       \left( \omega_{,\phi} \dot \phi^2 - f_{,\phi} + 2 V_{,\phi} \right) = 0,
   \label{BG}
\eea
where $H \equiv \dot a/a$.
The general formulation for handling the structure formation processes 
is presented in unified way in \cite{GGT-H,GGT-QFT,GGT-HN,GGT-CT}.
In the following we briefly summarize the formulation which will be
used in later sections.

Using a gauge invariant combination 
\bea
   \delta \phi_\varphi
       \equiv \delta \phi - {\dot \phi \over H} \varphi
       \equiv - {\dot \phi \over H} \varphi_{\delta \phi},
   \label{UCG-UFG}
\eea
the action valid to the second order in the scalar type perturbation
becomes (for derivation, see \cite{GGT-CT})
\bea
   \delta S
   &=& {1\over 2} \int a^3 Z \Bigg\{ \delta \dot \phi_\varphi^2
       - {1 \over a^2} \delta \phi_\varphi^{\;\; |\alpha}
       \delta \phi_{\varphi,\alpha}
   \nonumber \\
   & & \qquad + \;
       {1 \over a^3 Z} {H \over \dot \phi}
       \left[ a^3 Z \left( {\dot \phi \over H} \right)^\cdot \right]^\cdot
       \delta \phi_\varphi^2 \Bigg\} dt d^3 x.
   \label{Action-pert}
\eea
$\delta \phi_\varphi$ is a gauge invariant combination which is
$\delta \phi$ in the uniform-curvature gauge ($\varphi \equiv 0$).
The non-Einstein nature of the theory is present in a parameter $Z$ 
which is defined as
\bea
   Z (t) \equiv { \omega + {3 \dot F^2 \over 2 \dot \phi^2 F }
       \over \left( 1 + {\dot F \over 2 H F} \right)^2 }.
   \label{Z-def}
\eea
The equation of motion of $\delta \phi_\varphi$ becomes
\bea
   \delta \ddot \phi_\varphi
   &+& { (a^3 Z)^\cdot \over a^3 Z } \delta \dot \phi_\varphi
   \nonumber \\
   &-& \left\{ {1 \over a^2} \nabla^2 + {1\over a^3 Z} {H \over \dot \phi} 
       \left[ a^3 Z \left( {\dot \phi \over H} \right)^\cdot \right]^\cdot
       \right\} \delta \phi_\varphi = 0.
   \label{delta-phi-eq}
\eea
In terms of $\varphi_{\delta \phi}$ we have
\bea
   {1 \over a^3 Z} {H^2 \over \dot \phi^2}
       \left( a^3 Z {\dot \phi^2 \over H^2} \dot \varphi_{\delta \phi}
       \right)^\cdot - {1\over a^2} \nabla^2 \varphi_{\delta \phi} = 0.
   \label{varphi-eq}
\eea
The large and small scale asymptotic solutions are:
\bea
   \delta \phi_\varphi ({\bf x}, t)
   &=& - \;
       {\dot \phi \over H} \left[ C ({\bf x}) - D ({\bf x}) \int_0^t
       {1 \over a^3 Z} {H^2 \over \dot \phi^2} dt \right],
   \label{delta-phi-LS-sol} \\
   \delta \phi_\varphi ({\bf k}, \eta)
   &=&
       { 1 \over a \sqrt{2k} } \Big[ c_1 ({\bf k}) e^{i k \eta}
       + c_2 ({\bf k}) e^{-ik\eta} \Big] {1 \over \sqrt{Z} },
   \label{delta-phi-SS-sol}
\eea
where $C({\bf x})$ and $D({\bf x})$ are integration constants of the
growing and decaying mode, respectively.
[$D({\bf x})$ term is higher order in the large scale expansion
compared with the solutions in the other gauges; see \cite{GGT-HN}.]
At this point $c_1({\bf k})$ and $c_2({\bf k})$ are arbitrary
integration constants.

Equation (\ref{delta-phi-eq}) can be written as
\bea
   & & v^{\prime\prime} - \left( {z^{\prime\prime} \over z}
       + \nabla^2 \right) v = 0,
   \nonumber \\
   & & v \equiv \sqrt{Z} a \delta \phi_\varphi, \quad
       z \equiv \sqrt{Z} {a \dot \phi \over H}.
   \label{v-z-eq}
\eea
[The similar form in Einstein gravity and some types of generalized gravity
can be found in \cite{Mukhanov}.]
{\it When} we have $z^{\prime\prime} / z = n / \eta^2$ 
with $n = {\rm constant}$, Eq. (\ref{delta-phi-eq}) leads to a solution
\bea
   \delta \phi_{\varphi {\bf k}} (\eta)
   &=& {\sqrt{ \pi |\eta|} \over 2 a} \Big[ c_1 ({\bf k}) H_\nu^{(1)} (k|\eta|)
   \nonumber \\
   & & + \;
       c_2 ({\bf k}) H_\nu^{(2)} (k|\eta|) \Big] {1 \over \sqrt{Z}}, \quad
       \nu \equiv \sqrt{ n + {1\over 4} }.
   \label{delta-phi-k-sol}
\eea
Considering $\delta \phi_{\varphi {\bf k}} (\eta)$ as a mode function
of $\delta \hat \phi ({\bf x}, t)$ which is regarded as a quantum 
Heisenberg operator, the canonical quantization condition leads to the 
following normalization condition 
\bea
   & & | c_2 ({\bf k}) |^2 - | c_1 ({\bf k}) |^2 = 1.
   \label{normalization}
\eea
[The coefficients $c_i({\bf k})$'s in 
Eqs. (\ref{delta-phi-SS-sol},\ref{delta-phi-k-sol}) have phase differences.]
The power spectrum based on the vacuum expectation value is
\bea
   & & {\cal P}^{1/2}_{\delta \hat \phi_\varphi} ({\bf k}, \eta)
       = \sqrt{k^3 \over 2 \pi^2} | \delta \phi_{\varphi {\bf k}} |.
   \label{P-vac-def}
\eea
The corresponding power spectrum of $\hat \varphi_{\delta \phi}$ follows from
Eq. (\ref{UCG-UFG}) as
\bea
   & & {\cal P}^{1/2}_{\hat \varphi_{\delta \phi}} ({\bf k}, t)
       = {H \over \dot \phi} {\cal P}^{1/2}_{\delta \hat \phi_\varphi}
       ({\bf k}, t).
\eea

Using parameters
\bea
   & & \epsilon_1 \equiv {\dot H \over H^2}, \quad
       \epsilon_2 \equiv {\ddot \phi \over H \dot \phi}, \quad
       \epsilon_3 \equiv {1 \over 2} {\dot F \over H F}, \quad
       \epsilon_4 \equiv {1 \over 2} {\dot E \over H E},
   \nonumber \\
   & & E \equiv F \left( \omega + {3 \dot F^2 \over 2 \dot \phi^2 F} \right),
   \label{epsilons}
\eea
for $\dot \epsilon_i = 0$ we have
[for general expressions see Eq. (88) of \cite{GGT-HN}] 
\bea
   n 
   &=& { ( 1 - \epsilon_1 + \epsilon_2 - \epsilon_3 + \epsilon_4 )
       ( 2 + \epsilon_2 - \epsilon_3 + \epsilon_4 ) 
       \over ( 1 + \epsilon_1 )^2 }.
   \label{n}
\eea
{}For $\dot \epsilon_1 = 0$ we have
\bea
   & & \eta = - {1 \over aH} {1 \over 1 + \epsilon_1}.
\eea

%%%%%%%%%%%%%%%%%%%%%%%%%%%%%%%%%%%%%%%%%%%%%%%%%%%%%%%%%%%%%%
\section{Applications to Kinetic Inflations}
                                    \label{sec:KI}

%%%%%%%%%%%%%%%%%%%%%%%%%%%%%%%%%%%%%%%%%%%%%%%%%%%%%%%%%%%%%%
\subsection{Scalar-tensor Theory}
                                    \label{sec:ST}

A scalar-tensor theory is given by an action
\bea
   S = \int d^4 x \sqrt{-g} \left[ \phi R
       - \omega (\phi) { \phi^{;a} \phi_{,a} \over \phi} - V(\phi) \right],
   \label{ST-action}
\eea
which is a case of Eq. (\ref{GGT-action}) with
\bea
   f = 2 \phi R, \quad
       F \equiv {df \over dR} = 2 \phi, \quad
       \omega \rightarrow 2 {\omega (\phi) \over \phi}.
\eea
Equation (\ref{GGT-GFE}) becomes:
\bea
   & & R_{ab} = {1 \over \phi} \left[ \phi_{,a;b}
       + \omega {\phi_{,a} \phi_{,b} \over \phi}
       + {1 \over 2} \left( V + \Box \phi \right) g_{ab} \right],
   \nonumber \\
   & & \Box \phi + {1 \over 2 \omega + 3} \left( \omega_{,\phi}
       \phi^{;c} \phi_{,c} + 2 V - \phi V_{,\phi} \right) = 0.
\eea
Equation (\ref{BG}) becomes:
\bea
   & & H^2 = - H {\dot \phi \over \phi} + {\omega \over 6}
       {\dot \phi^2 \over \phi^2} + {V \over 6 \phi},
   \nonumber \\
   & & \ddot \phi + 3 H \dot \phi
       + {1 \over 2 \omega + 3} \left( \omega_{,\phi} \dot \phi^2 
       + \phi V_{,\phi} - 2 V \right) = 0.
   \label{BG-ST}
\eea

{\it Ignoring} the potential term, Eq. (\ref{BG}) leads to the
following solution \cite{Levin}:
\bea
   & & \dot \phi = C_1 { 1 \over a^3 \sqrt{1 + {2 \over 3} \omega} }, \quad
       \phi = \pm C_1 {1 \over a^2} \int {dt \over a}, 
   \nonumber \\
   & & H = {\dot \phi \over 2 \phi} \left( -1
       \pm \sqrt{ 1 + {2 \over 3} \omega } \right).
\eea
{\it If} we additionally have $\omega = {\rm constant}$ we can show:
\bea
   & & a \propto | t_0 - t |^{ - q }, \quad
       \phi \propto | t_0 - t |^{ 1 + 3q },
   \nonumber \\
   & & q \equiv - { 1 + \omega \mp \sqrt{1 + {2 \over 3} \omega}
       \over 4 + 3 \omega }.
   \label{BG-sol-ST}
\eea
A pole-like acceleration stage can be realized when $q>0$ which corresponds 
to the upper sign and $t_0 > t$.
In the following analyses for generality we will consider both signs. 

{}From Eq. (\ref{epsilons}) we have
\bea
   & & \epsilon_1 = {1 \over q}, \quad
       \epsilon_2 = - 3, \quad
       \epsilon_3 = - {1 + 3 q \over 2q}, \quad
       \epsilon_4 = 0.
\eea
Thus, from Eqs. (\ref{n},\ref{delta-phi-k-sol}) we have $n = - {1 \over 4}$ and 
\bea
   & & \varphi_{\delta \phi {\bf k}} (\eta)
       = {1 \over 4} \sqrt{\pi \over 3} 
       \left( \sqrt{ |\eta| \over a^2 \phi } \right)_1 
   \nonumber \\
   & & \qquad \times
       \Big[ c_1 ({\bf k}) H_0^{(1)} (k|\eta|)
       + c_2 ({\bf k}) H_0^{(2)} (k|\eta|) \Big].
   \label{varphi-k-sol-ST} 
\eea
The power-spectrum of fluctuating quantum field based on the vacuum expectation
value is presented in Eq. (\ref{P-vac-def}) for the metric 
fluctuations (coupled with the dilaton field) $\hat \varphi_{\delta \phi}$.
The corresponding power spectrums valid in general scales can be derived from
Eqs. (\ref{P-vac-def},\ref{varphi-k-sol-ST}).
In the large scale limit the power spectrum becomes:
\bea
   {\cal P}^{1/2}_{\hat \varphi_{\delta \phi}} ({\bf k}, \eta)
   &=& {1 \over \sqrt{3} } \left( \sqrt{|\eta| \over a^2 \phi} \right)_1
       \left( {k \over 2 \pi} \right)^{3/2} \ln{(k|\eta|)} 
   \nonumber \\
   & & \times 
       \Big| c_2 ({\bf k}) - c_1 ({\bf k}) \Big|.
   \label{P-C-QF-ST}
\eea

%%%%%%%%%%%%%%%%%%%%%%%%%%%%%%%%%%%%%%%%%%%%%%%%%%%%%%%%%%%%%%
\subsection{Induced Gravity Theory}
                                    \label{sec:Induced}

The induced gravity theory is given by
\bea
   & & S = \int d^4 x \sqrt{-g} \left[ {1 \over 2} \xi \phi^2 R
       - {1 \over 2} \phi^{;a} \phi_{,a} - V(\phi) \right],
   \label{Induced-action}
\eea
which is a case of Eq. (\ref{GGT-action}) with
\bea
   & & f = \xi \phi^2 R, \quad
       F = \xi \phi^2, \quad
       \omega = 1.
\eea
Equation (\ref{GGT-GFE}) becomes:
\bea
   & & R_{ab} = {1 \over \xi \phi^2} \Bigg[
       \phi_{,a} \phi_{,b} + g_{ab} V + \xi (\phi^2)_{,a;b}
       + {1 \over 2} \xi g_{ab} \Box \phi^2 \Bigg],
   \nonumber \\
   & & \Box \phi^2 
       = {2 \over 1 + 6 \xi} \left( \phi V_{,\phi} - 4 V \right).
\eea
Equation (\ref{BG}) becomes:
\bea
   & & H^2 + 2 H {\dot \phi \over \phi}
       - {1 \over 6 \xi} {\dot \phi^2 \over \phi^2}
       = {V \over 3 \xi \phi^2},
   \nonumber \\
   & & \ddot \phi + 3 H \dot \phi + {\dot \phi^2 \over \phi}
       = {1 \over 1 + 6 \xi} \left( {4 \over \phi} V
       - V_{,\phi} \right).
   \label{Induced-BG}
\eea

{\it Assuming} $V = 0$ Eq. (\ref{Induced-BG}) leads to the following 
solution \cite{Pollock}:
\bea
   & & a \propto | t_0 - t |^{-q}, \quad
       \phi \propto | t_0 - t |^{1 + 3q \over 2}, 
   \nonumber \\
   & & q \equiv - { 1 + 4 \xi \mp 4 \xi \sqrt{ 1 + {1 \over 6 \xi} }
       \over 3 + 16 \xi}.
   \label{BG-sol-Induced}
\eea
Cases with $q>0$, thus the upper sign, and $t_0 > t$
include the pole-like acceleration stage.
{}For generality, we will take both signs.

{}From Eq. (\ref{epsilons}) we have:
\bea
   & & \epsilon_1 = {1 \over q}, \quad
       \epsilon_2 = {1 - 3q \over 2q}, \quad
       \epsilon_3 = \epsilon_4 = - {1 + 3q \over 2q}.
\eea
Thus, from Eqs. (\ref{n},\ref{delta-phi-k-sol}) we have 
$n = - {1 \over 4}$ and
\bea
   & & \varphi_{\delta \phi {\bf k}} (\eta)
       = { \sqrt{\pi} \over 2 \sqrt{6\xi} }
       \left( { \sqrt{|\eta|} \over a \phi } \right)_1
   \nonumber \\
   & & \qquad \times
       \Big[ c_1 ({\bf k}) H_0^{(1)} (k|\eta|)
       + c_2 ({\bf k}) H_0^{(2)} (k|\eta|) \Big].
   \label{varphi-k-sol-induced}
\eea
The power spectrums in general scale is given by Eq. (\ref{P-vac-def}).
In the large scale limit we have:
\bea
   {\cal P}^{1/2}_{\hat \varphi_{\delta \phi}} ({\bf k}, \eta)
   &=& \sqrt{2 \over 3 \xi } 
       \left( { \sqrt{|\eta|} \over a \phi } \right)_1 
       \left( {k \over 2 \pi} \right)^{3/2} \ln{(k|\eta|)} 
   \nonumber \\
   & & \times
       \Big| c_2 ({\bf k}) - c_1 ({\bf k}) \Big|.
   \label{P-C-QF-Induced}
\eea

By the following changes the results in the scalar-tensor theory 
can be translated into the ones in the induced gravity 
\bea
   & & \phi \rightarrow {1 \over 2} \xi \phi^2, \quad
       \omega \rightarrow {1 \over 4 \xi}, \quad
       \delta \phi \rightarrow \xi \phi \delta \phi.
\eea

%%%%%%%%%%%%%%%%%%%%%%%%%%%%%%%%%%%%%%%%%%%%%%%%%%%%%%%%%%%%%%
\subsection{String Theory}
                                    \label{sec:string}

The low-energy effective action of string theory is given by
\cite{string-action}
\bea
   & & S = \int d^4 x \sqrt{-g} {1 \over 2} e^{-\phi}
       \Big[ R + \phi^{;a} \phi_{,a} - 2 V (\phi) \Big],
   \label{String-action}
\eea
which is a case of Eq. (\ref{GGT-action}) with
\bea
   & & f = e^{-\phi} R, \quad
       \omega  = - e^{-\phi}, \quad
       V \rightarrow e^{-\phi} V.
\eea
Equation (\ref{GGT-GFE}) becomes:
\bea
   & & R_{ab} = - \phi_{,a;b} - g_{ab} V_{,\phi}, 
   \nonumber \\
   & & \Box \phi = \phi^{;a} \phi_{,a} + 2 \left( V + V_{,\phi} \right).
\eea
Equation (\ref{BG}) becomes:
\bea
   & & H^2 = H \dot \phi - {1 \over 6} \dot \phi^2 + {1 \over 3} V,
   \nonumber \\
   & & \ddot \phi + 3 H \dot \phi - \dot \phi^2
       + 2 \left( V + V_{,\phi} \right) = 0.
   \label{BG-string}
\eea

{}For $V = 0$ Eq. (\ref{BG-string}) leads to the following solution
\bea
   & & a \propto | t_0 - t |^{\mp 1/\sqrt{3} }, \quad
       e^\phi \propto | t_0 - t |^{-1 \mp \sqrt{3} }.
   \label{BG-sol-string}
\eea
The upper sign with $t < t_0$ represents a pole-like inflation stage
which is called as a pre-big bang stage \cite{pre-big-bang}.
The corresponding studies of Sec. \ref{sec:ST} for the pre-big bang
scenario are presented in \cite{GGT-string}.
In the following we summarize the results considering both signs 
for generality.

{}From Eq. (\ref{epsilons}) we have
\bea
   & & \epsilon_1 = \epsilon_2 = \pm \sqrt{3}, \quad
       2 \epsilon_3 = \epsilon_4 = - \left( 3 \pm \sqrt{3} \right).
\eea
Thus, from Eqs. (\ref{n},\ref{delta-phi-k-sol}) we have $n = - {1 \over 4}$ and
\bea
   & & \varphi_{\delta \phi {\bf k}} (\eta)
       = - {1 \over 2} \sqrt{\pi \over 6} 
       \left( \sqrt{ |\eta| \over a^2 e^{-\phi} } \right)_1 
   \nonumber \\
   & & \qquad \times
       \Big[ c_1 ({\bf k}) H_0^{(1)} (k|\eta|)
       + c_2 ({\bf k}) H_0^{(2)} (k|\eta|) \Big].
   \label{varphi-k-sol-string}
\eea
The power spectrum valid in general scales can be derived from 
Eq. (\ref{P-vac-def}). 
In the large scale limit we have
\bea
   {\cal P}^{1/2}_{\hat \varphi_{\delta \phi}} ({\bf k}, \eta)
   &=& \sqrt{2 \over 3} \left( \sqrt{|\eta| \over a^2 e^{-\phi} } \right)_1
       \left( {k \over 2 \pi} \right)^{3/2} \ln{(k|\eta|)} 
   \nonumber \\
   & & \times 
       \Big| c_2 ({\bf k}) - c_1 ({\bf k}) \Big|.
   \label{P-C-QF-string}
\eea
The power spectrum of the dilaton field in Eq. (\ref{P-C-QF-string})
is derived in \cite{GGT-string}; 
authors in \cite{Brustein-etal} derived a similar spectrum in the context of 
the conformally transformed Einstein frame.

By following changes the results in the scalar-tensor theory 
can be translated into the ones in the pre-big bang scenario
\bea
   & & \omega \rightarrow -1, \quad
       \phi \rightarrow {1\over 2} e^{-\phi}, \quad
       V \rightarrow e^{-\phi} V,
   \nonumber \\
   & & \delta \phi \rightarrow - {1\over 2} e^{-\phi} \delta \phi.
   \label{ST-string}
\eea

%%%%%%%%%%%%%%%%%%%%%%%%%%%%%%%%%%%%%%%%%%%%%%%%%%%%%%%%%%%%%%
\subsection{Einstein Gravity with a Minimally Coupled Scalar Field}
                                   \label{sec:MSF-Kin}

The minimally coupled scalar field is a case of Eq. (\ref{GGT-action}) with
$f = R$ and $\omega = 1$.
Equation (\ref{GGT-GFE}) becomes:
\bea
   & & R_{ab} = 8 \pi G \left( \phi_{,a} \phi_{,b} + g_{ab} V \right), \quad
       \Box \phi - V_{,\phi} = 0.
   \label{Einstein-eq}
\eea
Equation (\ref{BG}) becomes:
\bea
   & & H^2 = {8 \pi G \over 3} \left( {\dot \phi^2 \over 2} + V \right), \quad
       \ddot \phi + 3 H \dot \phi + V_{,\phi} = 0.
   \label{BG-MSF}
\eea
By identifying $Z = F = \omega = 1$ equations in Sec. \ref{sec:Formulation} 
are valid for the minimally coupled scalar field.

{}For $V = 0$ Eq. (\ref{BG-MSF}) has the following solution
\bea
   a \propto | t_0 - t |^{1/3}, \quad
       \phi \propto \ln{|t_0 - t|}.
\eea
This expansion law corresponds to the ultra-relativistic limit with
an equation of state $p = \mu$.
Even in this case, driven by the pure kinetic term,
from Eqs. (\ref{epsilons},\ref{n}) we have $\epsilon_1 = \epsilon_2 = -3$
and $\epsilon_3 = \epsilon_4 = 0$, thus $n = - {1 \over 4}$.
{}For the mode function, Eq. (\ref{delta-phi-k-sol}) remains valid with 
$\nu = 0$ and $Z = 1$.
Thus, in the large scale limit, the power spectrum in Eq. (\ref{P-vac-def})
becomes
\bea
   {\cal P}^{1/2}_{\delta \hat \phi_\varphi} ({\bf k}, \eta)
   &=& 2 \left( { \sqrt{|\eta|} \over a } \right)_1
       \left( {k \over 2 \pi} \right)^{3/2} \ln{(k|\eta|)} 
   \nonumber \\
   & & \times 
       \Big| c_2 ({\bf k}) - c_1 ({\bf k}) \Big|.
   \label{P-C-QF-MSF}
\eea

%%%%%%%%%%%%%%%%%%%%%%%%%%%%%%%%%%%%%%%%%%%%%%%%%%%%%%%%%%%%%%
\subsection{Conventional Inflation with a Minimally Coupled Scalar Field}
                                   \label{sec:MSF}

Studies in the case of the ordinary inflation based on a minimally 
coupled scalar field are thoroughly presented in \cite{H-QFT};
compared with the previous works in \cite{Infl-spec-EXP,Infl-sp} the results in
\cite{H-QFT} are based on the uniform-curvature gauge or the equivalent 
gauge invariant combinations, $\delta \phi_\varphi$.
In the following, for comparison, we briefly rederive the results by 
reducing the general results in Sec. \ref{sec:Formulation}.
Equations (\ref{Einstein-eq},\ref{BG-MSF}) remain valid.
The cases where the background scale factor follows an exponential
or a power-law expansion in time correspond to $\dot \epsilon_i = 0$.
Thus, in both cases $n$ becomes constant and the solution in 
Eq. (\ref{delta-phi-k-sol}) applies.

{\it 1. Exponential expansion:} 
{}For $a \propto e^{Ht}$ with $H = {\rm constant}$ Eq. (\ref{BG-MSF}) 
has a solution with $V = {\rm constant}$ and $\dot \phi = 0$.
We have $\epsilon_i = 0$, thus $n = 2$ and $\nu = {3 \over 2}$.
{}From Eqs. (\ref{delta-phi-k-sol},\ref{P-vac-def}) we have the power spectrum 
valid in general scales and for the general vacuum state.
In the large scale limit we have
\bea
   {\cal P}^{1/2}_{\delta \hat \phi_\varphi} ({\bf k}, \eta)
   &=& {H \over 2 \pi} \Big| c_2 ({\bf k}) - c_1 ({\bf k}) \Big|.
   \label{P-EXP-S}
\eea
A choice of the adiabatic vacuum (known as the Bunch-Davies vacuum in 
de Sitter space) corresponds to $c_2({\bf k}) = 1$ and $c_1({\bf k}) = 0$.
The spectrum in Eq. (\ref{P-EXP-S}) with the adiabatic vacuum was
found in \cite{Infl-spec-EXP}

{\it 2. Power-law expansion:} 
{}For $a \propto t^p$ with $p = {\rm constant} (> 1)$ Eq. (\ref{BG-MSF})
has a solution with $\dot \phi = \sqrt{2p}/t$ and 
$V = p (3p -1)/t^2 \propto e^{-\sqrt{2/p} \phi}$, \cite{Lucchin-etal}.
In this case we have $\epsilon_1 = \epsilon_2 = - 1/p$, 
thus $\nu = \nu_g = {3 p - 1 \over 2(p - 1)}$.
The general power spectrum follows from 
Eqs. (\ref{delta-phi-k-sol},\ref{P-vac-def}). 
In the large scale limit we have
\bea
   {\cal P}^{1/2}_{\delta \hat \phi_\varphi} ({\bf k}, \eta)
   &=& H { \Gamma(\nu) \over \pi^{3/2} } { p - 1 \over p }
       \left( {2 \over k |\eta|} \right)^{\nu - 3/2}
   \nonumber \\
   & & \times
       \Big| c_2 ({\bf k}) - c_1 ({\bf k}) \Big|.
   \label{P-POW-S}
\eea
In the limit of $p \rightarrow \infty$ Eq. (\ref{P-POW-S}) reproduces 
Eq. (\ref{P-EXP-S}).
The spectrum in Eq. (\ref{P-POW-S}) with an adiabatic vacuum was first
rigorously derived in \cite{Infl-spec-POW}.

%%%%%%%%%%%%%%%%%%%%%%%%%%%%%%%%%%%%%%%%%%%%%%%%%%%%%%%%%%%%%%
\section{Density Spectrums}
                                   \label{sec:Spectrums}

In the previous section we have derived quantum fluctuations based on the
vacuum expectation value of the fluctuating quantum fields of
the scalar type perturbations which include the fluctuating dilaton field
and the fluctuating metric.
The derived mode functions and the power spectrums in 
Sec. \ref{sec:ST}-\ref{sec:string} are valid for the {\it general} 
expansion stages considered in 
Eqs. (\ref{BG-sol-ST},\ref{BG-sol-Induced},\ref{BG-sol-string}), respectively.
Notice that the expansion stages included in 
Eqs. (\ref{BG-sol-ST},\ref{BG-sol-Induced},\ref{BG-sol-string})
are not necessarily acceleratory.
The reason why the power spectrums are insensitive to the specifics 
of the expansion law remains to be explained.
The expansion stages include the pole-like accelerated expansion stages.
Although constructing the specific models for a successful inflation 
remains to be 
seen, in the following we will assume a scenario that the pole-like 
acceleration stages provide the inflation era in the early universe.

We consider an evolution scenario of the presently observable patch of the
universe and the structures in it as follows.
The first assumption concerns quantum generation processes:
The observationally relevant structures in the present universe
are {\it supposed} to exit the local horizon, thus becoming the superhorizon 
scale later on, {\it during} the kinetic inflation stage supported by 
one of the generalized gravity theories.
This first assumption allows us to consider the quantum fluctuations 
based on the vacuum expectation value as possible natural seeds for the later
evolution into the large scale structures.
The second assumption concerns the classical evolution processes:
the underlying gravity governing the dynamics of our patch of the universe 
is {\it supposed} to transit from one of the generalized gravity to 
Einstein gravity {\it while} the relevant scales we are considering 
were in the superhorizon scale.
Since the evolution of linear structures in superhorizon scales is kinematic 
in nature, we can conveniently handle the evolution using the conserved 
quantities.
{}For the scalar type perturbations the perturbed three space curvature
variable $\varphi$ in the uniform-field gauge (equivalently, the
comoving gauge in Einstein's gravity), i.e. $\varphi_{\delta \phi}$,
is the one which is temporally conserved.
{}From Eqs. (\ref{UCG-UFG},\ref{delta-phi-LS-sol}), ignoring the decaying mode
which is higher order in the large scale limit, we have 
\bea
   & & \varphi_{\delta \phi} ({\bf x}, t) = C ({\bf x}).
   \label{varphi-conserv}
\eea
We emphasize that Eq. (\ref{varphi-conserv}) is valid considering general 
changes in $V(\phi)$, $\omega (\phi)$ and $f(\phi,R)$ in the 
generalized gravity \cite{GGT-H}, and in the equation of state $p = p(\mu)$ 
in the fluid era \cite{H-IF}. 
This reflects the kinematic nature of the evolution in the superhorizon scale.
Thus, this second assumption makes the handling of the classical evolution
processes easy.
Equation (\ref{varphi-conserv}) remains valid virtually in all scales
(more precisely, larger than Jeans scale) while the universe is in 
the matter dominated era, \cite{H-IF}.
{}From the classical power spectrum of $C({\bf x})$,
we can derive the power spectrums
for the rest of the scalar type perturbations, like, fluctuations in
the density, potential, and velocity fields and also the directional 
fluctuations in the cosmic microwave background photons.

In the large scale limit the classical power spectrum of 
$C({\bf x})$ based on the spatial averaging becomes
\bea
   {\cal P}^{1/2}_{C} ({\bf k})
   &=& {\cal P}^{1/2}_{\varphi_{\delta \phi}} ({\bf k}, \eta)
   \nonumber \\
   &=& {\cal P}^{1/2}_{\hat \varphi_{\delta \phi}} ({\bf k}, \eta)
       \times \sqrt{ Q (k) },
   \label{P-C}
\eea
where in the first step we used Eqs. (\ref{UCG-UFG},\ref{delta-phi-LS-sol}) 
neglecting the decaying mode, and in the second step we adopted an ansatz
on matching the classical fluctuations with the quantum fluctuations.
$Q(k)$ is a classicalization factor which may take into account of
possible effects from the classicalization process \cite{Hu}.
{}From ${\cal P}_C$ we have the power spectrums of the classical fields like
the density, velocity, gravitational potential, scalar contribution to the
temperature anisotropy of the CMBR, etc. 
In Eq. (\ref{P-C}) the right hand side should be evaluated while the scale 
is in the large scale limit during the inflation era.
The power spectrums of the quantum fluctuations, 
${\cal P}^{1/2}_{\hat \varphi_{\delta \phi}} ({\bf k}, \eta)$ 
during various kinetic inflation stages are derived Sec. \ref{sec:KI}.

In the second horizon crossing epoch where the matter is dominated we have 
\bea
   {\cal P}^{1/2}_{\delta} (k,t_{\rm HC})
       = {2 \over 5} {\cal P}^{1/2}_{C} (k)
       = { (a H)^2 \over \sqrt{2} \pi } k^{-1/2} |\delta_{\bf k} (t)|.
   \label{HC-power-spectrum}
\eea
Conventionally we take $|\delta_{\bf k} (t)|^2 \equiv A(t) k^n$
where $n$ is a spectral index. 
Ignoring the classicalization factor ($Q \equiv 1$),
the vacuum dependence (thus, taking $c_2 \equiv 1$ and $c_1 \equiv 0$), 
and the logarithmic dependence on $k$ we have a generically
tilted spectrum with $n = 4$ for the scenarios considered
in \S \ref{sec:ST}-\ref{sec:MSF-Kin}.
The observation of the large scale cosmic structures and particularly the
large scale anisotropy of the CMBR indicate $n \simeq 1$.
The spectrum with $n \simeq 1$ is the Zeldovich spectrum which is a
natural outcome of the near exponential inflation considered
in Eqs. (\ref{P-EXP-S},\ref{P-POW-S}).
The observationally relevant temperature fluctuations of the CMBR 
in different direction, $\delta T$, can be related with the fluctuating 
metric as ${\delta T \over T} = {1 \over 5} C$.
Using
${\delta T \over T} (\theta,\phi) = \sum_{lm} a_{lm} Y_{lm} (\theta,\phi)$,
we have
\bea
   & & a_l^2 \equiv \langle |a_{lm}|^2 \rangle
       = {4 \pi \over 25} \int_0^\infty {1\over k}
       {\cal P}_C (k) j_l^2 (kx) d k,
   \label{a_l}
\eea
where $x = 2/H_0$.

Thus, the scenario we have considered in this section 
with one of the quantum fluctuations in \S \ref{sec:ST}-\ref{sec:MSF-Kin}
as the seed fails to produce the observationally relevant structures.
This implies that the pole-like inflation stages are not suitable
for the seed generating mechanism for the observed large scale structures.

%%%%%%%%%%%%%%%%%%%%%%%%%%%%%%%%%%%%%%%%%%%%%%%%%%%%%%%%%%%%%%
\section{Discussion}
                                    \label{sec:Discussion}

In Sec. \ref{sec:KI} we have shown that the power spectrum
${\cal P}_{\delta \hat \phi_\varphi} \propto k^{3}$
is a robust prediction from all the expansion stages driven by the
kinetic (more precisely, potential-less) parts in the gravity.
If one accepts this quantum fluctuations as the seed for the later 
evolution into the large scale classical structures through
the inflation mechanism, as shown in Sec. \ref{sec:Spectrums}, 
it leads to a completely different spectrum for the large scale 
structures compared with the observed ones.  
We may call it a ``structure problem'' of the kinetic inflation scenarios.
Together with the ``graceful exit problem'', which generically appears
in the kinetic type inflation scenarios, this can be accepted as 
another negative news for constructing inflation models based on 
a pole-like expansion stage in the generalized gravity theories.

We can think of some ways out of this generic structure problem in the
kinetic inflation.
Firstly, the power spectrums of the quantum fluctuations depend on 
the vacuum choice which is almost an arbitrary function of $k$
with a constraint in Eq. (\ref{normalization}); the choice of the vacuum
state needs physical consideration.
Secondly, the classicalization process of the quantum fluctuations
can possibly lead to a modification factor $Q(k)$ which in general may
depend on the wave number $k$; however, the effects may arise from considering 
the nonlinear field effect which goes beyond the linear treatment considered 
in this paper \cite{Hu}.
Finally, probably the most reasonable approach
would be to accept it as a problem for making the observed part of 
the large scale structures.
In such a case, the observed part of the large scale structure can possibly
arise from other seed generating mechanisms, like defects.
This implies that kinetic inflation based on the above gravity theories
is not appropriate for the seed generating mechanism for the currently 
observable patch of the universe.
However, one remarkable point of the pole-like acceleration stage is that
as the expansion proceeds the curvature term grows because
$R \sim H^2 \sim |t_0 - t|^{-q - 2}$.
Thus, as $t$ approaches to $t_0$, the curvature term diverges
and the quantum effect may become significant again in the later stage
of the acceleration era.
In such a case, the classical actions used in Sec \ref{sec:KI}
should be modified by the quantum correction terms.
Recently, it was suggested that the graceful exit problem in the
kinetic inflation based on the string gravity can be 
resolved by the one loop quantum correction effect \cite{Rey}.
In such a modified scenario, the observationally relevant scales may
exit the horizon during the quantum era.
Deriving the generated quantum fluctuations in such modified gravity
sectors may lead to completely different outcomes and require considering 
the action which is more general than the one in Eq. (\ref{GGT-action}).
This is currently an important open question especially in the
context of the recently popular string theory.

The case of a decoupled gravitational wave can be treated in a similar manner.
Each of two polarization states of the gravitational wave behaves 
similarly to the scalar type perturbation and we can analogously derive 
the exact forms of the power spectrums of the quantum fluctuations.
The power spectrums show similar dependences on $k$ to the power spectrums 
for the scalar type perturbations presented in Sec. \ref{sec:KI}.
Results for the gravitational wave will be presented separately.

%%%%%%%%%%%%%%%%%%%%%%%%%%%%%%%%%%%%%%%%%%%%%%%%%%%%%%%%%%%%%%
\section*{Acknowledgments}

This work was supported by the KOSEF, Grants No. 95-0702-04-01-3 and 
No. 961-0203-013-1, and through the SRC program of SNU-CTP.

%%%%%%%%%%%%%%%%%%%%%%%%%%%%%%%%%%%%%%%%%%%%%%%%%%%%%%%%%%%%%%
\section*{Appendix}

This appendix contains the equations and the general asymptotic solutions 
for the scalar type perturbations in three types of gravity theories 
considered in Sec. \ref{sec:ST}-\ref{sec:string}.

{\it 1. Scalar-tensor theory:}
It is convenient to have:
\bea
   & & {\dot \phi \over H \phi} = - {1 + 3 q \over q}, \quad
       {1 + q \over 1 + 3 q} = \sqrt{ 1 + {2 \over 3} \omega },
   \nonumber \\
   & & Z = 12 \left( {q \over 1 + 3 q} \right)^2 {1 \over \phi},
   \\
   & & \eta = - {1 \over 1 + q} {t_0 - t \over a} 
       = - {q \over 1 + q} {1 \over aH}.
   \nonumber
\eea
{}For the scalar mode Eqs. (\ref{UCG-UFG}-\ref{delta-phi-SS-sol}) 
become:
\bea
   & & \delta \phi_\varphi = {1 + 3 q \over q} \phi \varphi_{\delta \phi},
   \label{UCG-UFG-ST} \\
   & & \delta S = {1 \over 2} \int a^3 Z \Bigg[ \delta \dot \phi^2_\varphi
       - {1 \over a^2} \delta \phi_\varphi^{\;\;|\alpha}
       \delta \phi_{\varphi,\alpha}
   \nonumber \\
   & & \qquad
       - \left( {1 + 3 q \over t_0 - t} \right)^2 \delta \phi_\varphi^2 \Bigg] 
       dt d^3 x,
   \label{Action-pert-ST} \\
   & & \delta \ddot \phi_\varphi + {1 + 6 q \over t_0 - t}
       \delta \dot \phi_\varphi
       + \left[ \left( {1 + 3 q \over t_0 - t} \right)^2
       - {1 \over a^2} \nabla^2 \right] \delta \phi_\varphi = 0,
   \nonumber \\
   \label{delta-phi-eq-ST} \\
   & & \ddot \varphi_{\delta \phi}
       - {1\over t_0 - t} \dot \varphi_{\delta\phi}
       - {1 \over a^2} \nabla^2 \varphi_{\delta \phi} = 0,
   \label{varphi-eq-ST} \\
   & & \varphi_{\delta \phi} = C + {D \over 12} 
       \left( {t_0 - t \over a^3 \phi} \right)_1 \ln{(1 - t/t_0)},
   \label{varphi-LS-sol-ST} \\
   & & \varphi_{\delta \phi} = {1 \over 2 \sqrt{6}}
       \left( \sqrt{ |\eta| \over a^2 \phi } \right)_1
       { 1 \over \sqrt{k|\eta|} } 
       \left[ c_1 e^{ik\eta} + c_2 e^{-ik\eta} \right].
   \label{varphi-SS-sol-ST}
\eea

{\it 2. Induced gravity:}
It is convenient to have:
\bea
   & & {\dot \phi \over H \phi} = - {1 + 3 q \over 2 q}, \quad
       {1 + q \over 1 + 3q} = \sqrt{ 1 + {1\over 6 \xi} },
   \nonumber \\
   & & Z = 6 \xi \left( {2q \over 1 + 3 q} \right)^2,
   \\
   & & \eta = - {1 \over 1 + q} {t_0 - t \over a}
       = - {q \over 1 + q} {1 \over a H}.
   \nonumber
\eea
{}For the scalar mode Eqs. (\ref{UCG-UFG}-\ref{delta-phi-SS-sol}) become:
\bea
   & & \delta \phi_\varphi = {1 + 3q \over 2q}
       \phi \varphi_{\delta \phi},
   \\
   & & \delta S = {1 \over 2} \int a^3 Z \Bigg[ \delta \dot \phi^2_\varphi
       - {1 \over a^2} \delta \phi_\varphi^{\;\;|\alpha}
       \delta \phi_{\varphi,\alpha}
   \nonumber \\
   & & \qquad \qquad
       - \left( {1 + 3 q \over 2} {1 \over t_0 - t} \right)^2
       \delta \phi_\varphi^2 \Bigg] dt d^3 x,
   \\
   & & \delta \ddot \phi_\varphi
       + {3 q \over t_0 - t} \delta \dot \phi_\varphi
       + \left[ \left( {1 + 3 q \over 2 (t_0 - t)} \right)^2
       - {1 \over a^2} \nabla^2 \right] \delta \phi_\varphi = 0,
   \nonumber \\
   \\
   & & \ddot \varphi_{\delta \phi}
       - {1\over t_0 - t} \dot \varphi_{\delta\phi}
       - {1 \over a^2} \nabla^2 \varphi_{\delta \phi} = 0,
   \\
   & & \varphi_{\delta \phi} = C + {D \over 6 \xi} 
       \left( {t_0 - t \over a^3 \phi^2} \right)_1
       \ln{(1 - t/t_0)},
   \\
   & & \varphi_{\delta \phi} = {1 \over 2 \sqrt{3 \xi}}
       \left( { \sqrt{|\eta|} \over a \phi } \right)_1
       { 1 \over \sqrt{k|\eta|} } 
       \left[ c_1 e^{ik\eta} + c_2 e^{-ik\eta} \right].
\eea

{\it 3. String theory:}
It is convenient to have
\bea
   & & {\dot \phi \over H} = 3 \pm \sqrt{3}, \quad
       Z = \left( 2 \mp \sqrt{3} \right) e^{-\phi}.
   \nonumber \\
   & & \eta = - {3 \mp \sqrt{3} \over 2} {t_0 - t \over a} 
       = - {1 \pm \sqrt{3} \over 2} {1 \over aH}.
\eea
{}For the scalar mode Eqs. (\ref{UCG-UFG}-\ref{delta-phi-SS-sol}) become:
\bea
   & & \delta \phi_\varphi = - \left( 3 \pm \sqrt{3} \right)
       \varphi_{\delta \phi},
   \label{UCG-UFG-string} \\
   & & \delta S = {1 \over 2} \int a^3 Z \Bigg( \delta \dot \phi^2_\varphi
       - {1 \over a^2} \delta \phi_\varphi^{\;\;|\alpha}
       \delta \phi_{\varphi,\alpha} \Bigg) dt d^3 x,
   \label{Action-pert-string} \\
   & & \ddot \varphi_{\delta \phi}
       - {1\over t_0 - t} \dot \varphi_{\delta\phi}
       - {1 \over a^2} \nabla^2 \varphi_{\delta \phi} = 0,
   \label{varphi-eq-string} \\
   & & \varphi_{\delta \phi} = C + {D \over 6} 
       \left( { t_0 - t \over a^3 e^{-\phi} } \right)_1 \ln{(1 - t/t_0)},
   \label{varphi-LS-sol-string} \\
   & & \varphi_{\delta \phi} = - {1 \over 2 \sqrt{3}}
       \left( \sqrt{ |\eta| \over a^2 e^{-\phi} } \right)_1
       { 1 \over \sqrt{k|\eta|} } 
       \left[ c_1 e^{ik\eta} + c_2 e^{-ik\eta} \right].
   \nonumber \\
   \label{varphi-SS-sol-string}
\eea

%%%%%%%%%%%%%%%%%%%%%%%%%%%%%%%%%%%%%%%%%%%%%%%%%%%%%%%%%%%%%%

%%%%%%%%%%%%%%%%%%%%%%%%%%%%%%%%%%%%%%%%%%%%%%%%%%%%%%%%%%%%%%

\begin{references}
\bibitem{Inflation}
         A. Linde, {\it Particle Physics and Inflationary Cosmology},
            (Harwood Academic Publishers: New York, 1990);
         E. W. Kolb and M. S. Turner, {\it The Early Universe},
               (Addison-Wesley Publishing Company: Seoul, 1994).
\bibitem{Pollock}
         M. D. Pollock and D. Sahdev, Phys. Lett. B {\bf 222}, 12 (1989).
\bibitem{pre-big-bang}
         G. Veneziano, Phys. Lett. B {\bf 265}, 287 (1991);
         M. Gasperini and G. Veneziano, Astroparticle Phys. {\bf 1}, 317 (1993).
\bibitem{Levin} 
         J. J. Levin and K. Freese, Nucl. Phys. B {\bf 421}, 635 (1994);
         J. J. Levin, Phys. Rev. D {\bf 51}, 462 (1995);
         Phys. Rev. D {\bf 51}, 1536 (1995);
         Phys. Lett. B {\bf 343}, 69 (1995).
\bibitem{GGT-H}
         J. Hwang, Phys. Rev. D {\bf 53}, 762 (1996).
\bibitem{GGT-QFT}
         J. Hwang, Report. No. gr-qc/9607059 (unpublished).
\bibitem{GGT-HN}
         J. Hwang and H. Noh, Phys. Rev. D {\bf 54}, 1460 (1996).
\bibitem{GGT-string}
         J. Hwang, Report. No. hep-th/9608041 (unpublished).
\bibitem{GGT-CT}
         J. Hwang, Report. No. gr-qc/9605024 (unpublished).
\bibitem{Mukhanov}
         V. F. Mukhanov, Soviet Phys. JETP {\bf 68}, 1297 (1988);
         V. F. Mukhanov, H. A. Feldman and R. H. Brandenberger,
               Physics Reports {\bf 215}, 203 (1992).
\bibitem{string-action}
         C. G. Callan, D. Friedan, E. J. Martinec and M. J. Perry,
               Nucl. Phys. B {\bf 262}, 593 (1985).
\bibitem{Brustein-etal}
         R. Brustein, M. Gasperini, M. Giovannini, V. Mukhanov and
            G. Veneziano, Phys. Rev. D {\bf 51}, 6744 (1995).
\bibitem{H-QFT}
         J. Hwang, Phys. Rev. D {\bf 48}, 3544 (1993);
         Astrophys. J. {\bf 427}, 542 (1994);
         Class. Quantum Grav. {\bf 11}, 2305 (1994).
\bibitem{Infl-spec-EXP}
         A. Guth and S. Pi, Phys. Rev. Lett. {\bf 49}, 1110 (1982);
         S. W. Hawking, Phys. Lett. B {\bf 115}, 339 (1982);
         A. A. Starobinsky, Phys. Lett. B {\bf 117}, 175 (1982);
         J. M. Bardeen, P. S. Steinhardt and M. S. Turner,
               Phys. Rev. D {\bf 28}, 679 (1983).
\bibitem{Infl-sp}
         R. H. Brandenberger and R. Kahn, Phys. Rev. D {\bf 29}, 2172 (1984);
         J. J. Halliwell and S. W. Hawking, Phys. Rev. D {\bf 31}, 1777 (1985);
         D. H. Lyth, Phys. Rev. D {\bf 31}, 1792 (1985);
         V. F. Mukhanov, JETP Lett. {\bf 41}, 493 (1985);
         M. Sasaki, Prog. Theor. Phys. {\bf 76}, 1036 (1986);
         V. F. Mukhanov, Soviet Phys. JETP {\bf 68}, 1297 (1988).
\bibitem{Infl-spec-POW}
         D. H. Lyth and E. D. Stewart, Phys. Lett. B {\bf 274}, 168 (1992).
\bibitem{COBE}
         A. R. Liddle and D. H. Lyth, Phys. Rep. {\bf 231}, 1 (1993);
         M. White, D. Scott and J. Silk, Ann. Rev. Astron. Astrophys. 
            {\bf 32}, 319 (1994).
\bibitem{Hu}
         E. Calzetta and B. L. Hu, Phys. Rev. D {\bf 52}, 6770 (1995).
\bibitem{Lucchin-etal}
         F. Lucchin and S. Matarrese, Phys. Rev. D {\bf 32}, 1316 (1985).
\bibitem{H-IF}
         J. Hwang, Astrophys. J. {\bf 415}, 486 (1993);
         Astrophys. J. {\bf 427}, 533 (1994).
\bibitem{Rey}
         S.-J. Rey, Phys. Rev. Lett. {\bf 77}, 1929 (1996).
\end{references}
\end{document}